# Correctness, confidence, and context:

## Framing software assurance in the AI age

Narrative for keynote presentation at FSE 2026, July 2026

Mary Shaw
Software & Societal Systems Dept
Carnegie Mellon University
Pittsburgh PA USA
mary.shaw@cs.cmu.edu

## ABSTRACT

Software engineering has a complicated relationship with "correctness". We recognize the challenges of full formal rigor as well as many required properties beyond functional correctness. Although we satisfice in practice, we are still stuck in the mindset that we could reason our way to correctness, if only we had enough information. Unfortunately for our hopes of formal rigor, both our expectations of real-world systems and our implementation decisions are shaped by unspoken knowledge that is personal, subjective, qualitative, and largely unavailable for this analysis.

Generative AI has introduced a new dimension to assurance: its foundation is statistical rather than formal. Traditional software engineering establishes confidence through rigorous reasoning, domain knowledge and expert judgment. In contrast, generative AI results are sophisticated predictions, "probably approximately correct" [36]. This inherently limits assurances about the results to probabilistic assertions.

Further, the nuances that guide human judgment are often tacit or implicit, not explicitly described. This knowledge casts only scant shadows into the digital record, so that critical source of knowledge is only faintly represented in AI models.

We have many approaches for developing assurances that a software system does what it's expected to do, though most of them focus on code specifications rather than system requirements, let alone the system's fitness for its purpose. We have failed to develop a systematic understanding of the relative merits of the various approaches to assurance. I hope that generative AI will finally force us to tackle this.

To that end, I will challenge us to think systematically about our assurance techniques, especially the role of hidden context and the challenges of AI. We need ways to make informed, reasoned choices about cost-effective combinations of approaches to developing confidence in our systems.

We call ourselves software engineers. Let's act like engineers.

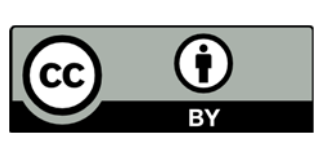

## 1. Software Engineering and Correctness

Software engineering has a complicated relation with "correctness". On the one hand, our roots are in the symbolic, logical, precise reasoning of computer science, and we have a deep-seated urge to prove our software is correct. On the other hand, we recognize that this is usually impractical or even impossible for a variety of reasons, ranging from intellectual complexity, to the variety of properties we care about, to incomplete knowledge, to engineering pragmatics about cost-effectiveness.

The tension between formality and pragmatics often arises over context—the extent to which a piece of software should be evaluated in isolation or in its complete context of use. Much of this context is tacit or implicit—folklore, culture, domain expertise, and non-documented intent—but we tend to ignore these knowledge sources when we define development processes. Practically, we should seek confidence that our software is fit for its intended purpose. This, of course, requires us to address the purpose of the software: how good software for that task *needs* to be and how to manage the risk that it will fail [12].

The tension is heightened by modern AI, which couples great power with great uncertainty, all sitting on a statistical predictive basis rather than a rigorous mathematical one. It produces results that are, in Leslie Valiant's description [36] "probably approximately correct." SE is, of course, familiar with uncertainty and nondeterminism. However, the statistical basis of LLMs is of a different kind from that of classical systems in which nondeterminism arises, for example, from the physical properties of embedded systems or from variations in input data. Reasoning about this difference in statistical foundations is one of the major challenges AI now presents to SE.

Further, much of the tacit and implicit context of intended use is inaccessible to AI training sets, as is much of the design intent and tradeoff analysis that lies behind the implementation. This information is critical to the success of the systems, but it is often undocumented or even unrecognized. Only through the scant traces it leaves in the accessible records—its *digital shadow*—can it affect the AI outcomes.

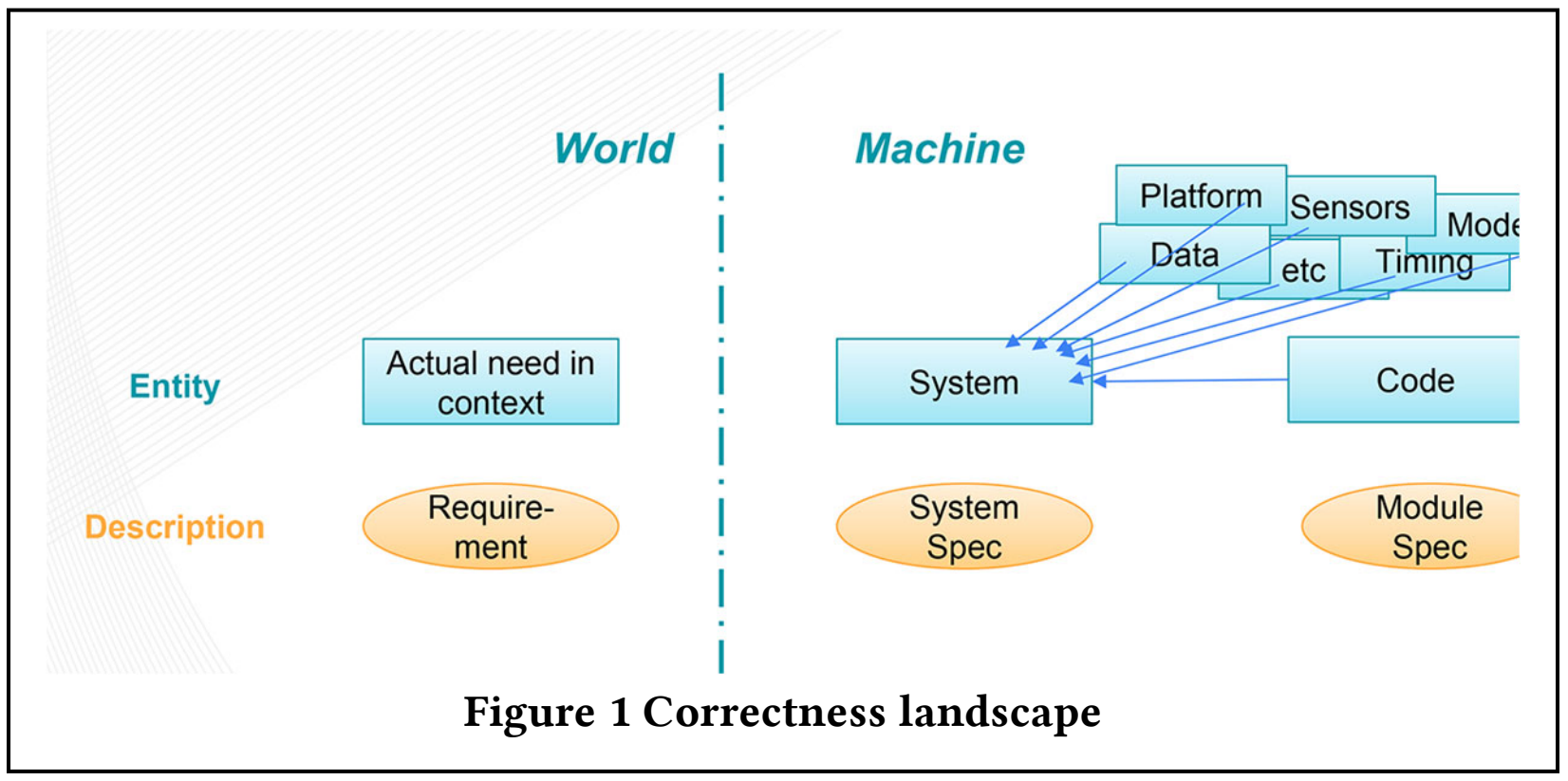

**Figure 1 Correctness landscape**

I've been pondering this tension for some time, hoping that modern AI will finally force SE to be more systematic. I'll lay out some of the tensions about correctness within SE, discuss the kinds of tacit and implicit information that are often neglected, identify some inherent problems with AI, and propose a framework for reasoning about assurances. This will recognize that any software project needs to satisfy only a subset of the large set of properties that are sometimes desirable. It will also offer a way to choose from among the many techniques available to us for evaluating the properties, placing techniques for AI and AI-generated elements on the same footing as the rest.

This is not the first time I've raised the question of whether AI will finally force SE to be more systematic about correctness.

The first time was in my keynote for the History of Programming Languages Conference in 2021 [28], where I challenged our long-standing mindset that we create software to solve well-specified problems and prove that it is correct. We do, of course, but only sometimes. More often we satisfice [32]—we work until we have a good enough result for the problem at hand. At HOPL I said, "It may be time to rethink what we mean by correctness for a complex software system, especially an embedded system."

The second time was in my Onward! essay with Kang [13] where we addressed the responsibilities of software engineering that go beyond just coding to cover all of software. We discussed the ways that SE has regularly adapted to new technology, especially new AI technology over the years. We said, "SE needs to rethink its concept of correctness; generative AI may finally force SE to do this".

## 2. The Landscape of "Correctness"

Software is written to serve human needs, so I'll begin with Jackson's distinction [11] between the world of human needs and the machine that's serving those needs. On the "world" side of that distinction we have the actual need, in its full context, and a proxy for that need in the form of a requirement. On the "machine" side we have the system and the software components incorporated in the system, each with its own specification. This is depicted in Figure 1.

This view is insufficient, however. Although SE tends to regard a system as simply a composition of code modules, real systems actually integrate components of many types. Even for simple standalone systems, data is as important as code and less often validated.

An especially poignant example occurred on July 19 2024, when Crowdstrike distributed an updated configuration file for their security software. A flaw in the updated data caused widespread outages on 8.5 million systems running Windows software [4]. Many of these systems were critical to business and public services around the world, most notably healthcare, banking, and airlines, with estimated damages over $5 billion dollars [5]. The flaw was identified within hours, and a repair was distributed. However, many of the crashed systems had to be restarted manually, so the disruption from the outages persisted for days. Their initial analysis was, in essence, "we validated the last code release in February; it has been fine for months. All we did last night was release data (a configuration file)"[3].

More broadly, software development relies on software development ecosystems that include platforms, frameworks and databases as well as languages. These cluster around classes of applications. Further, the languages are often domain-specific languages, not general-purpose languages, and most have not been rigorously designed. Figure 2 draws on a StackOverflow survey to show the ecosystem of tools used in that community for web development and Microsoft technologies [39].

This richness of structure is shown in Figure 3, which extends Figure 1 by adding the class of properties that the system developer should establish. Three of these are the usual correspondences between the entities and their descriptions: in the language of the world (fitness for purpose), and in the language of the machine (conformance of the system to its specification and correctness—often functional correctness—of the code in a module). An additional correspondence is required, that the system specification (in the language of the machine) is a faithful rendition of the requirement (in the language of the world). This last one is particularly challenging, because humans are notoriously bad at expressing full descriptions of requirements in natural language, and defining this correspondence requires understanding both the world and the machine. It is also essential, because it is the locus of the knowledge about how concepts in the world are realized in the implementation of the machine.

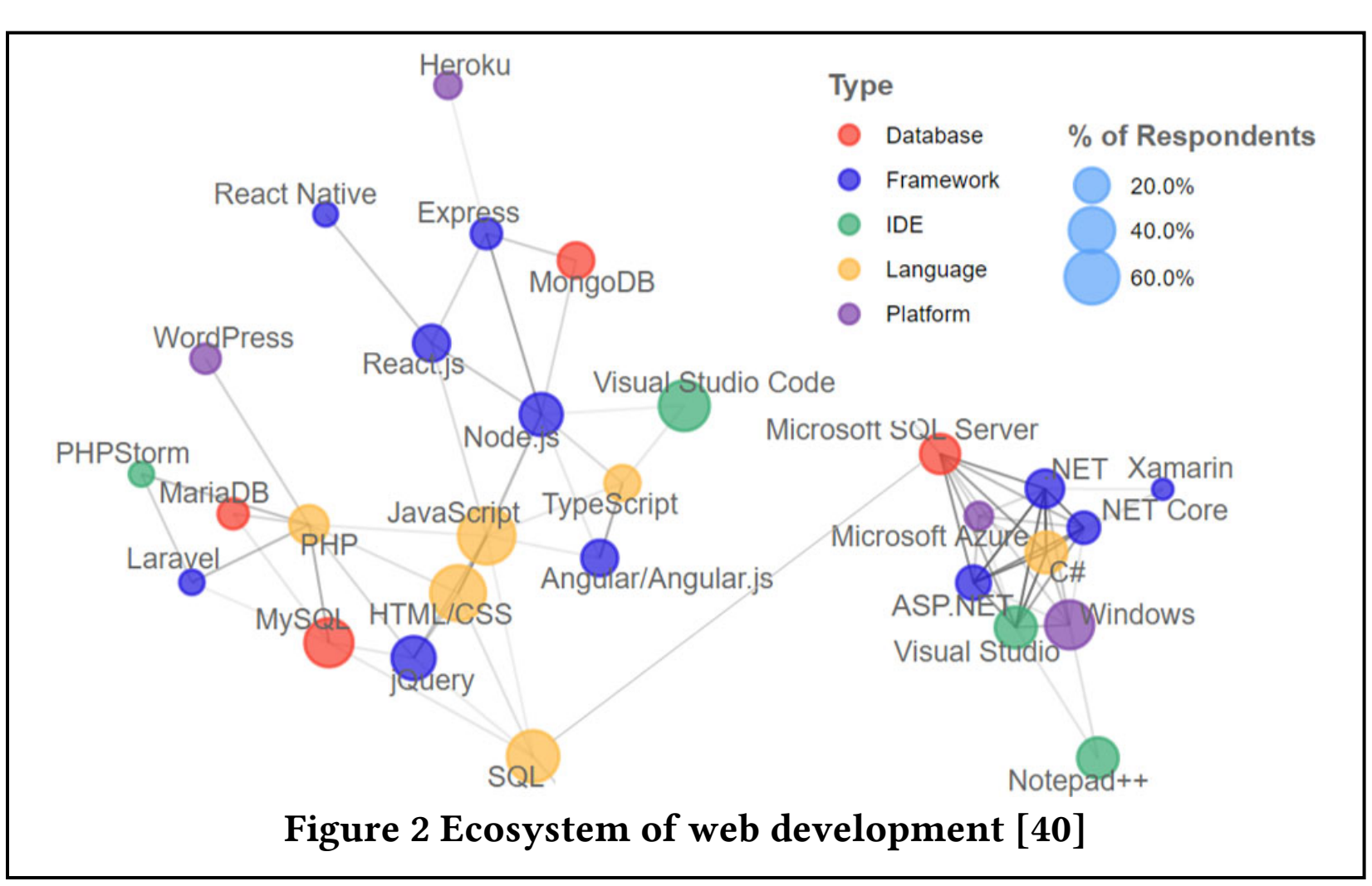

**Figure 2 Ecosystem of web development [40]**

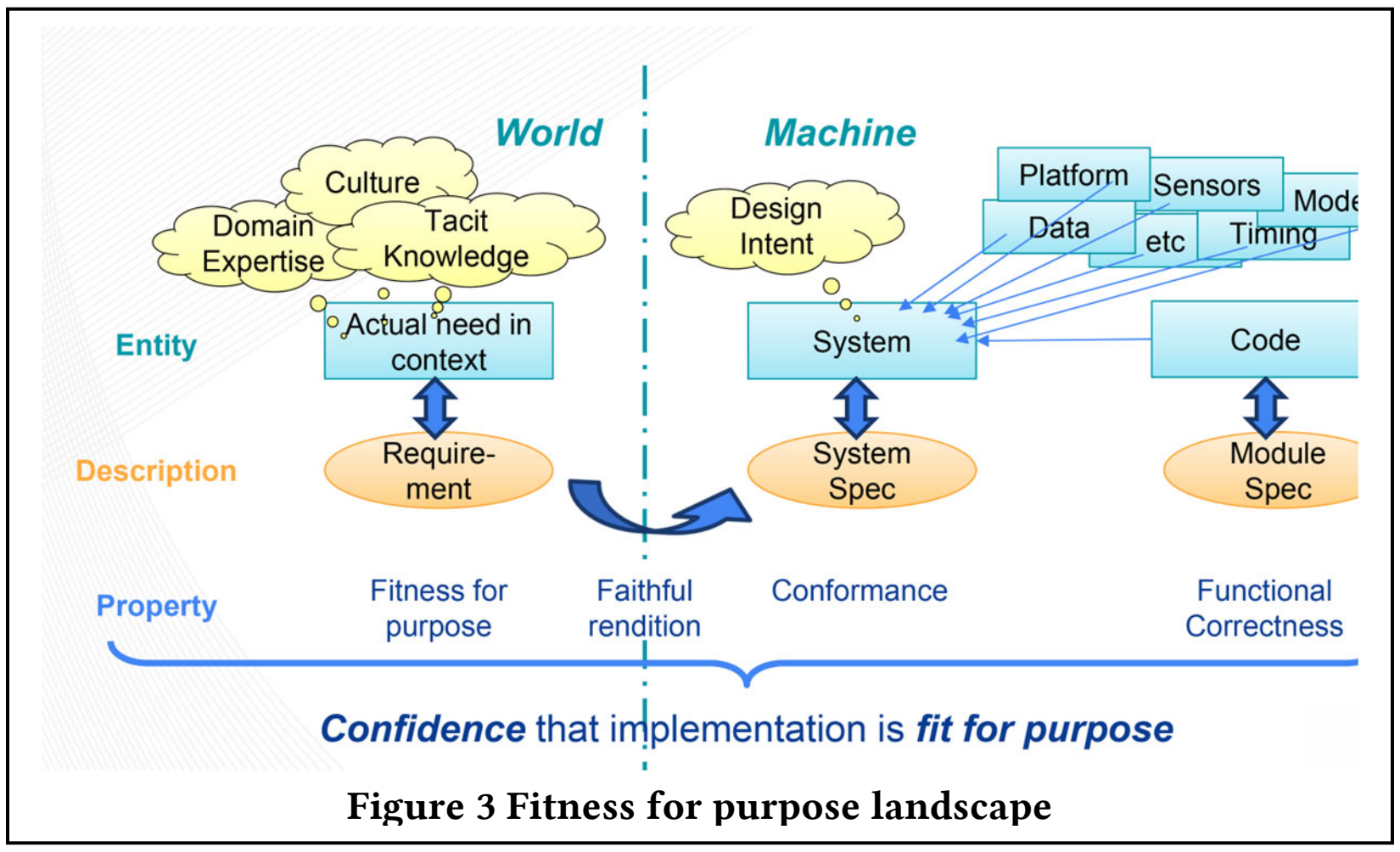

**Figure 3 Fitness for purpose landscape**

There is a further, quite significant consideration: context. Figure 3 also adds this to the landscape. On the machine side, the designer's intent is often not captured, or only imperfectly captured. Yet this is the intellectual backbone of the integrity of the system, and it is often accessible only as oral history within a project.

**Don't assume that formal specifications can capture everything you care about.**[1]

More seriously, the world has an enormous amount of tacit or implicit knowledge that does not become explicit in the requirement. The most obvious is domain knowledge—the knowledge of a domain expert beyond what's written in handbooks and standards. More nuanced are the opinions of various stakeholders who are affected by the project. In describing "wicked problems", Rittel and Webber discussed the challenges of precisely defining projects embedded in social settings, including reconciling disparate views of the problem and acceptable solutions and indeed the unwillingness of some stakeholders to voice opinions that challenged social norms [22]. Still more subtle are cultural norms, deeply seated worldviews that differ from the worldviews of dominant cultures, that members of the culture may not be able to articulate in ways the software developer understands, and that can make or break a project.

**Problem setting is as important as problem solving**

**Satisfice—fitness for the task at hand is often a better goal than formally verified correctness**

In this landscape we can see how the ideals of our formal mindset may not be achievable, both because of incomplete information and because of the effort of implementation and verification. In practice, then, we accept imperfect and incomplete results. We *satisfice*, working until we find a sufficient solution and not expending resources on finding a perfect one. Acknowledging that we may not achieve formal correctness does not absolve us of responsibility. Rather, it imposes a new obligation to decide, for each application, what level of confidence is required for each of its important properties.

Practical systems, then, must satisfy the requirement that the results must be good enough—we need *sufficient correctness*. Of course, we sometimes need the most nearly perfect software we can achieve. But much more often the cost of achieving that level of perfection is not justified, because incorrect answers or failures can be handled (or even ignored) at less cost than preventing them. Figure 4 illustrates some aspects of this tradeoff. It lays out a number of applications (as used in their customary contexts) along two axes: the consequences of failure and the degree of automation, or the diminishing likelihood that a human observer will notice a budding problem and intervene [41]. This separates the realm of "good enough" from the realm of "correctness." The figure also provides a warning about mission creep: successful products tend to accumulate new responsibilities. As the degree of automation grows and the consequences of failure increase, the threshold of confidence should similarly rise.

Lampson advises us to do end-to-end checks on system designs [15]. The end-to-end expectation of this correctness landscape, as shown in Figure 3, is confidence that the implementation (in the machine) is fit for its purpose (in the world). Thus this is more properly called the "fitness for purpose" landscape. Crucially, expected fitness depends on context. That context determines which properties matter and what level of confidence is needed. Some of the context will be given in the requirements, but much of it is implicit or tacit: it might be design decisions drawing on expert intuition, expectations that the client didn't think to mention, or cultural values that shape expectations. In doing the end-to-end check, knowing the structure of the software is much less important than knowing the reasons it was structured that way.

## 3. Implicit and Tacit Context

Software engineering has a complicated relation with documentation. Documentation is notoriously hard to create and quick to become obsolete. Yet there is no other record of the judgment calls designers make about the overall concept of the software or about tradeoffs that led to current decisions—but could change if the operating environment were to change. Similarly, the client's expectations often include criteria that the client didn't think were significant, or even values that are culturally embedded so strongly that the client isn't even aware of them.

Fitness for purpose often depends on this implicit or tacit context. Implicit context is, in principle, inferable from other infor-mation. Indeed, the Unblocked tool aspires to reconstruct context of a software project by piecing together fragments of project discussions such as notes and conversation—but it needs to have access to all the internal discussions [35]. Tacit context is undocumented and often not even consciously articulated. It may be the product of habit, it may arise from cultural norms, it may have dee-seated religious roots.

[1] All the boxed callouts are drawn from the "Creative Nudges" of [29]

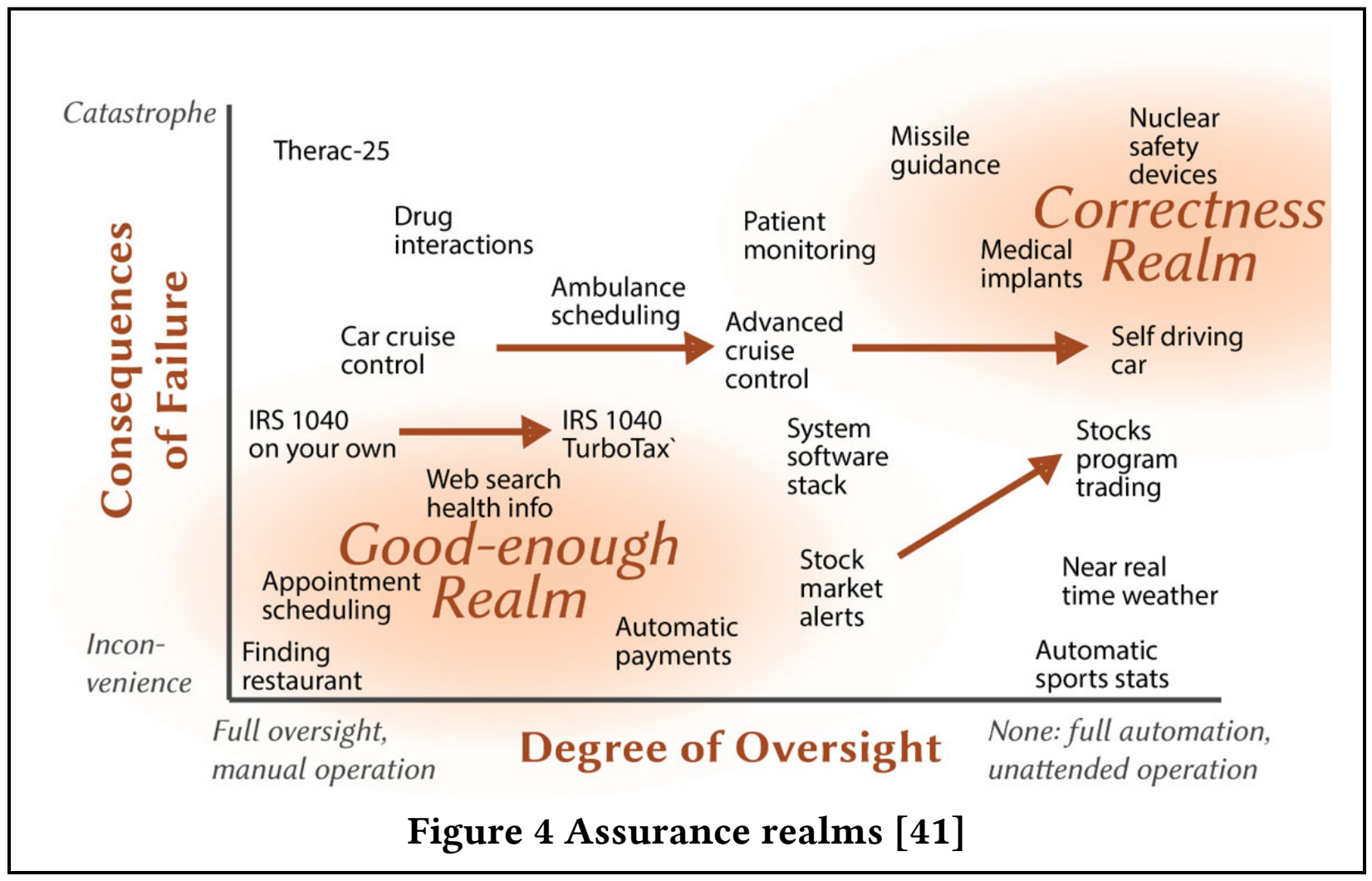

**Figure 4 Assurance realms [41]**

Understanding a system, both the needs in the world and the implementation in the machine, requires understanding of the reasons for the requirements and specifications. Without this, reasons for the requirement will fade and the project will accumulate cognitive and intend debt; the design integrity of the system will degrade and incur technical debt [34].

To show the importance of context that isn't explicitly part of the requirements or specifications, I'll offer a few examples, both in the language of the machine and the language of the world.

### Implicit machine context: conflicting assumptions

Garlan and colleagues reported in 1995 on their experience developing a system from components that were designed to be reusable. They encountered serious integration problems [7]. For example, the message protocols of the components were mismatched, and several components thought they "owned" the main event loop. The authors of the components had made good-faith efforts to document the components, but they were not aware that these attributes of the components would conflict with other reusable components.

This kind of problem persists. In his ICSA26 keynote, Binns reported on architectural challenges for software that measures wafers in a semiconductor manufacturing facility [1]. His component collected data intended for internal use to improve performance. When he changed the data he collected because the process changed, downstream clients complained—it turned out that other users had become dependent on the unpublished data that was intended to be private. Even bug fixing could create problems; if downstream clients had discovered the bug and compensated for it, then fixing the bug caused over-compensation.

It turns out to be harder than it looks to develop and enforce complete interfaces.

### Tacit machine context: the "why" resides in peoples' heads

The trend of delegating software development to AI tools is running into limitations. AI has been successful in many situations where the new software is very similar to existing software. We're seeing the limitations of this, however: AI is more successful in routine engineering than in innovative engineering. Two recent reports show the challenges of automating tasks that depend on tacit knowledge about software structure and design[2].

Larson expresses concern that software development is an oral tradition—that not much is written down and most of that is obsolete [16]. The important bits are reasons for decisions. This, he says, is fine for societies passing down culture from generation to generation, but the turnover in software engineering is so rapid that there's a steady drain on domain knowledge, with risk of increasing technical debt.

Steghöfer and Borg are concerned that the use of vibe coding for highly complex, interconnected systems creates maintainability issues because the AI tools assume that enough relevant information is contained in the code [33]. To the contrary, they say, domain knowledge, design rationales, risk management and the like will never be in the code, because they are captured in layers of abstraction above the code. These abstractions are the lifeblood of software engineering, and creating meaningful abstractions of complex systems is uniquely human.

**Specify "what", not "how"**

The 2026 International Conference on Software Architecture (ICSA2026)[10] had several papers on using AI to discover architecture from code. They were largely successful at recovering the module structure, interactions, and dependencies, but they were largely unsuccessful at recovering the abstractions that led to those structures.

This problem pervades engineering of all kinds. Ford Motor recently discovered that trying to replace people with AI led to loss of critical knowledge. They are trying to hire back senior engineers who have developed this knowledge through years of experience [19].

I've seen this in person. A number of years ago I visited the plant of a chemical processing company. They had just installed a centralized control room where data from all over that plant could be monitored conveniently. They were finding that this did not provide as much control over the plant as they expected, because experienced operators anticipated problems from subtle sensory signals as they walked through the plant—a new small here, a change of sound there, a whiff of oil somewhere else. If the operators were in the control room, though, they were not walking through the plant and these tacit indicators were not accessible to them. When I visited, they were looking for ways to reconcile the advantage of the control room with the need for subtle feedback.

[2] Calling generative AI's errors "hallucinations" is appalling. It anthropomorphizes the software, and it spins actual errors as somehow being idiosyncratic quirks of the system even when they're objectively incorrect. Moreover, WebMD's advice about hallucinations is "If you or a loved one has hallucinations, go see a doctor." https://www.webmd.com/schizophrenia/what-are-hallucinations

Losing track of the big picture—the concept of the system, the reasoning behind design and implementation decisions—creates cognitive and intent debt as well as technical debt [34].

### Implicit world context: disciplinary lenses

One way to characterize disciplines is by the kinds of phenomena they study, the questions they value, and the methods they recognize for investigating these questions. This disciplinary mindset shapes the way practitioners address problems.[3]

Schön gives an extended example that illustrates this well. He describes attempts to solve a problem of malnutrition in the small village of Cali, Colombia [25]. Eight experts from different disciplines offered 8 different solutions.

Despite general consensus about the urgency of the problem, each group approached the problem from a different perspective. Nutritionists tried to choose the best diet selected from available foods at current prices. But total agricultural production wouldn't sustain the population. Agronomists saw the problem as one of agricultural productivity and tried to expand agricultural land and change the crop mix to increase protein. But in some villages malnourished children remained malnourished because of diarrhea caused by intestinal parasites. Public health specialists therefore focused on water quality, hygiene and sanitation. Economists attributed poor sanitation and low-protein diets to poverty. They devised training programs to raise income levels. Other observers noted that the fertile lowlands were devoted to raising crops for export while the peasants made a bare living from small farms on the mountainsides. They argued that a redistribution of land and wealth was needed. Researchers with a historical perspective saw that 20 years earlier Columbia produced enough food for its population, but population had grown faster than production thanks to successful public health interventions.

**Import ideas from other fields; adopt and adapt their problems or techniuqes**

The experts viewed the problem through the lenses of their own disciplines. Not only did they offer approaches based in the techniques of their own disciplines, but even their statement of the problem and the desired solution were framed in their customary disciplinary vocabularies. Schön follows this with a discussion of how a systems view could integrate the work of the individual specialties.

Default mindsets shape interpretation and even language.

### Tacit world context: ritual and causality

During World War II, Allied armed forces on Melanesian islands were resupplied by air, bringing material abundance to the islands. After the military departed at the end of the war, these resupply flights stopped. However, the indigenous population acted out the behavior of the military—marching in formation, raising flags, lighting the runway, even building wooden models of airplanes and bamboo control towers—as rituals in the belief that these were the actions that attracted the airplanes with the valuable goods. They had mistaken the incidental activities of the troops for the unobserved logistic and scheduling activities, and they had made sense of the phenomena by believing that the cargo must have been a gift from the gods, that the foreigners had simply discovered the rituals that unlocked the treasures, and that if they faithfully reproduced the rituals the cargo would return [38]. These beliefs and rituals are often called "cargo cults".

Many cultures have their own ways of sense-making that do not align with the ways of modern technology

### Rich and tacit world context: common objects associations

It can be hard to interpret even common words and concepts if they are used in many ways, including with tacit associations. This leads to a two-fold challenge: selecting which of the many associations are appropriate in any particular setting, and discerning the deeper associations behind the chosen ones.

Consider the sunflower. It is a plant, taxonomically Helianthus annuus, a large annual forb in the daisy family Asteraceae [37]. In plain language, it is a tall flowering plant whose flowers feature a wide central disk framed by yellow petals. But it can also be, explicitly,

- the source of sunflower oil, the 4th most-used vegetable oil in the world, with implications for production and commerce
- food for birds or livestock, often a byproduct of oil production, so interacting with that production
- a seed for snacking, with nutritional association that may depend n user's dietary preferences
- the national symbol of Ukraine, with deep associations of resistance and hope
- a filter for extracting heavy metals and nucleotides through the roots
- an ornamental flower
- a Zuni treatment for snakebite
- an element of Iroquois and Onondaga creation stories
- the sun god in Aztec and Inca traditions
- the state flower of Kansas, USA
- a symbol of long life, good fortune, and prosperity in China
- a favorite subject of Vincent Van Gogh
- a running example of concepts of computer science[20]

These associations are well documented, though their associations are nuanced. But there are also personal associations that have no digital presence and therefore are invisible to LLMs,

- topic of a gardening discussion with a Ukrainian neighbor
- depicted on my souvenir of a recent ICSA talk in Amsterdam, a scarf purchased at a local open-air market (at least this was tacit, up until I posted this document online!)
- myriad other things that none of us knows about

[3] I once watched two physicists plan a picnic. It was all about forces and masses.

If a requirement refers to sunflowers, which of these associations is important? Is your client a food processor, a politician, an artist, an Iroquois? What other associations does the client have, and why doesn't the client realize they're important?

### Explicit and implicit world and machine context: cultural literacy

Hayakawa devotes a chapter to "presymbolic" use of language [8]. It's common to express feelings through the loudness and tone of voice of the expression rather than the words themselves: If a car is about to hit someone, the alarm in your voice conveys more meaning than your choice of "hey" or "look out". Beyond that, though, social chatter is often about establishing trust, not actually exchanging information about the weather, the family, or the outcome of last night's game. Ritual utterances like this reinforce social cohesion.

Hirsch showed how effective communication in the world depends on a large body of shared knowledge and shared values, not simply on fluency in the language [9]. He concentrated on the body of knowledge shared within the United States: the Constitution, the Bible, Greek/roman mythology, the western canon of literature.[4] People from other cultures have other bodies of cultural knowledge.

Navigating a culture effectively also depends on tacit knowledge about things like the degree to which rules of public behavior must actually be followed, appropriate personal space and eye contact, the way to navigate public transportation, which type of store carries what you're looking for[5], and so on.

Similar cultural knowledge is associated with the use of software tools. It's not enough to know the syntax of a programming language or the operation of a tool. There are often associated design obligations [30]—folklore about good and bad ways to use the tools.

### Software engineering involves nuance and judgment

I call out these examples because they are too rarely recognized in software engineering discussions of "correctness" or confidence in fitness for purpose. There's a wealth of similar examples [31]. Just as the part of iceberg visible above water is much smaller than the part of the iceberg beneath the surface, the code of a system is much smaller than the other context and knowledge needed to evolve and maintain the system.

These examples reinforce guidance from the cultural canon of software engineering, and I call on your cultural knowledge of software engineering to need nothing more than brief reminders.

**Problems embedded in society are not often solved using ordinary methods of analysis**

Rittel and Webber's "wicked problems" paper showed that problems embedded in society are ill-defined, have many competing objectives and hidden agendas, and no clear success criterion [22]. They are not often solved with customary methods of analysis

**Distinguish essential from accidental complexity**

Brooks called out the difference between the "essence" and the "accident" of design: fashioning complex conceptual constructs is the essence; accidental tasks arise in representing these constructs [2].

Schön described design as a "reflective conversation with the materials of the situation" [25]. The materials' voice is silent; it is through interacting with the materials that the designer comes to understand the design problem and its solution.

**Reuse proven solutions when you can**

Software engineering is much more than creating code. By analogy to the definition of classical engineering, it is the branch of computer science that creates practical, cost-effective solutions to computing and information processing problems, by applying the best-systematized knowledge available, developing software systems in the service of mankind [26].

**Less than 10% of the code goes to the overt function of the system; the rest goes to housekeeping**

Engineering often distinguishes between innovative and routine design; the former finds novel solutions to unfamiliar problems; the latter builds variants on familiar solutions [26]. Even in a novel application, the majority of the code is devoted to routine housekeeping. This may account in part for AI's success in code generation. However, AI success at routine tasks has led to elevated expectations about complex innovative tasks.

## 4. AI Challenges and Inherent Limitations

I turn now to the relation between AI tools and achieving confidence that a system is fit for its purpose. I see three inherent limitations of AI for this purpose. First, a great deal of tacit knowledge is not available in a form that makes it accessible for LLM training sets. Second attempting to go directly from natural language problem descriptions to running code short-cuts the world-machine mapping. Third, the statistical basis of machine learning is unsound; it does not permit rigorous reasoning, and SE has not yet come to understand how to manage this difference.

**Automate things that are well-understood, tedious, and error-prone when left to humans**

AI brings great capability for carrying out routine tasks. Applying it to novel or complex tasks, however, carries risk. We should be judicious about applying it: we should automate things that are well-understood

[4] Yes, this is biased; but everyone's culture is. I recall when an Indian colleague was criticized for not knowing the Roman gods and myths. He replied "I do know the gods and myths. But mine are different from yours".

[5] On a bicycle camping trip in Finland, I needed to purchase alcohol fuel for my stove. I knew that the Finnish name was denaturoitu sprii with brand names Sinol and Marinol. But it took me a day and a half to figure out that it was not sold in hardware stores, camping stores, grocery stores, or pharmacies—what seemed like the obvious places. The place to buy it was at gasoline stations, but not just any gasoline station, it had to be a staffed station, not a fully automated station.

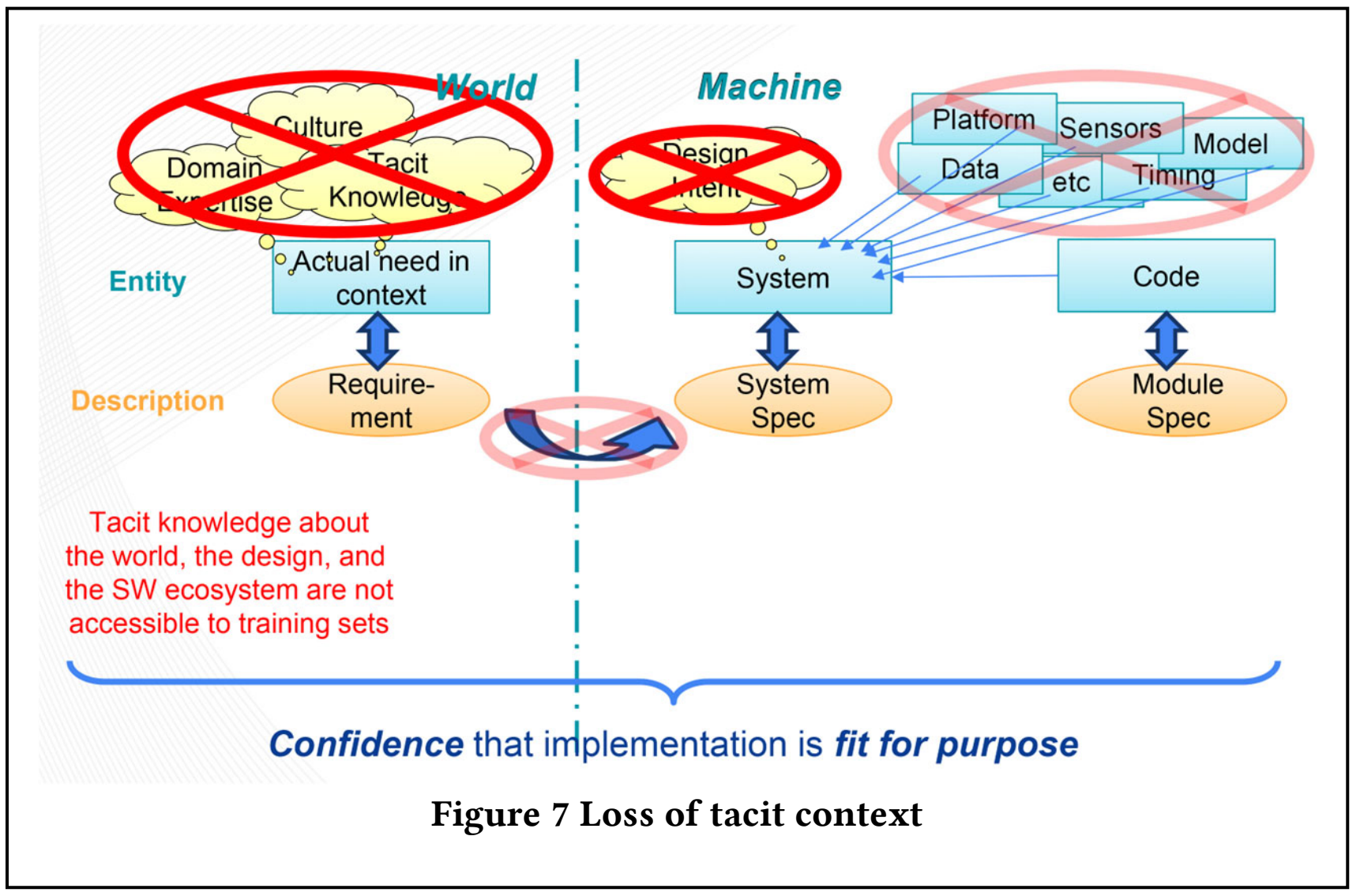

**Figure 7 Loss of tacit context**

and that are tedious and error-prone in the hands of humans. Humans should handle other things.

I'll illustrate each of these in the landscape of fitness for purpose of Figure 3.

### Inaccessible tacit knowledge

When important contextual information is not accessible to AI tools, it cannot affect the results from the tools. Figure 7 shows how inaccessibility of tacit knowledge impoverishes the correctness landscape. The lost information is marked in red. The loss of tacit and implicit knowledge about the world obscures domain expertise and nuances about the actual need and may cause the requirement to miss its target. The loss of design intent hides the abstractions of the design and increases reliance on the project's oral history; expertise about the development ecosystem such as the design obligations of various parts of the ecosystem may also be hidden. Implicit knowledge may also be inaccessible because the chain of reasoning to recognize it is too long. Finally, the faithful rendition of the requirement into the system specification depends on the lost tacit knowledge, so the mapping from the world to the machine is at risk.

In Plato's Allegory of the Cave [21], subjects see only shadows of objects, never the actual objects. The allegory shows how people mistake sensory perceptions for reality. The much simpler version of this in Figure 6a illustrates a well-known visual puzzle: what object makes a square shadow when viewed from the front, a triangular shadow when viewed from the side, and a round shadow when viewed from the top? The figure depicts a simple solid shape, but it gives no further information about the object casting the shadow. When I first encountered this puzzle, the object offered as the solution was a paper drinking cup that folded flat and was sometimes dispensed at water fountains; that is, the shape was an empty shell, and the solution relied on cultural knowledge that has now become obsolete. One version of the cup appears in Figure 6b [43]. Thus the shadows of the object provide only partial information about even the physical form of the object.

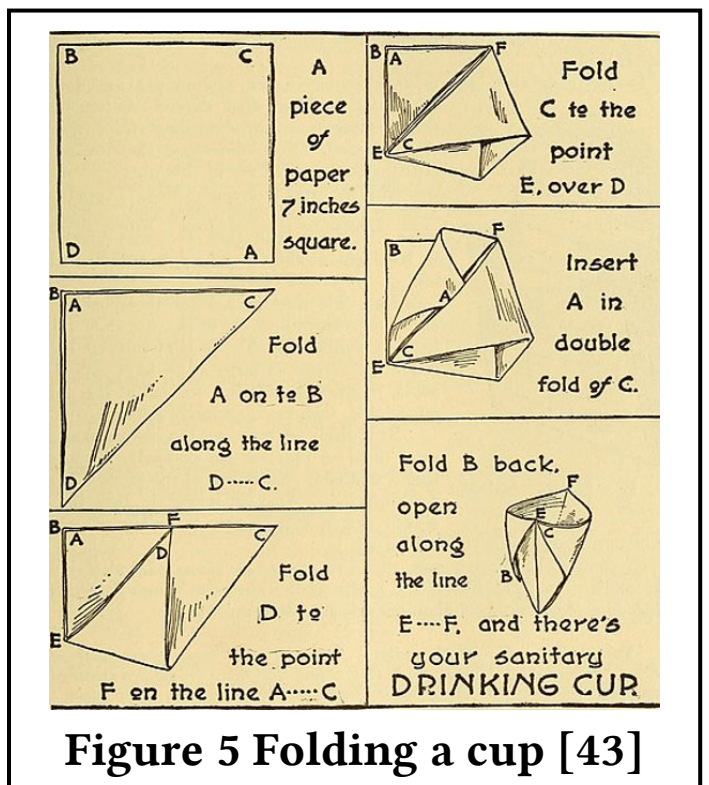

**Figure 5 Folding a cup [43]**

The cup in that figure has additional context: It must hold water, at least briefly, and it is disposable. It can be folded from a single sheet of paper as shown in Figure 5. This particular version has further context: it was (re?)discovered over a century ago to improve hygiene in a tuberculosis asylum.

This illustrates an inherent limitation of LLMs. They are trained on the *digital shadows*[6] of reality; they can only incorporate information that's captured in the records on which they are trained. This is true of both knowledge in the world and knowledge in the machine. Just as the shadow projections of the block and the cup omitted important information, the digital shadows of real phenomena omit tacit and implicit knowledge.

**Figure 6 Shadow projections of (a) a block and (b) a cup [42][43]**

### Shortcutting world-machine mapping

There has been much excitement and optimism recently about generative AI producing running code from prompts written in natural language—vibe programming. The optimism extends to the idea that vernacular programmers as well as highly-trained computing professionals won't have to worry about the details. This ignores design decisions about representations and architectures that affect security, maintainability, the ability to integrate with other components, and other quality attributes.

[6] This usage of "digital shadows' is consistent with industrial usage, which refers to a digital representation of a physical system. Both share the limitation that they only include information that is digitally available; neither has complete knowledge.

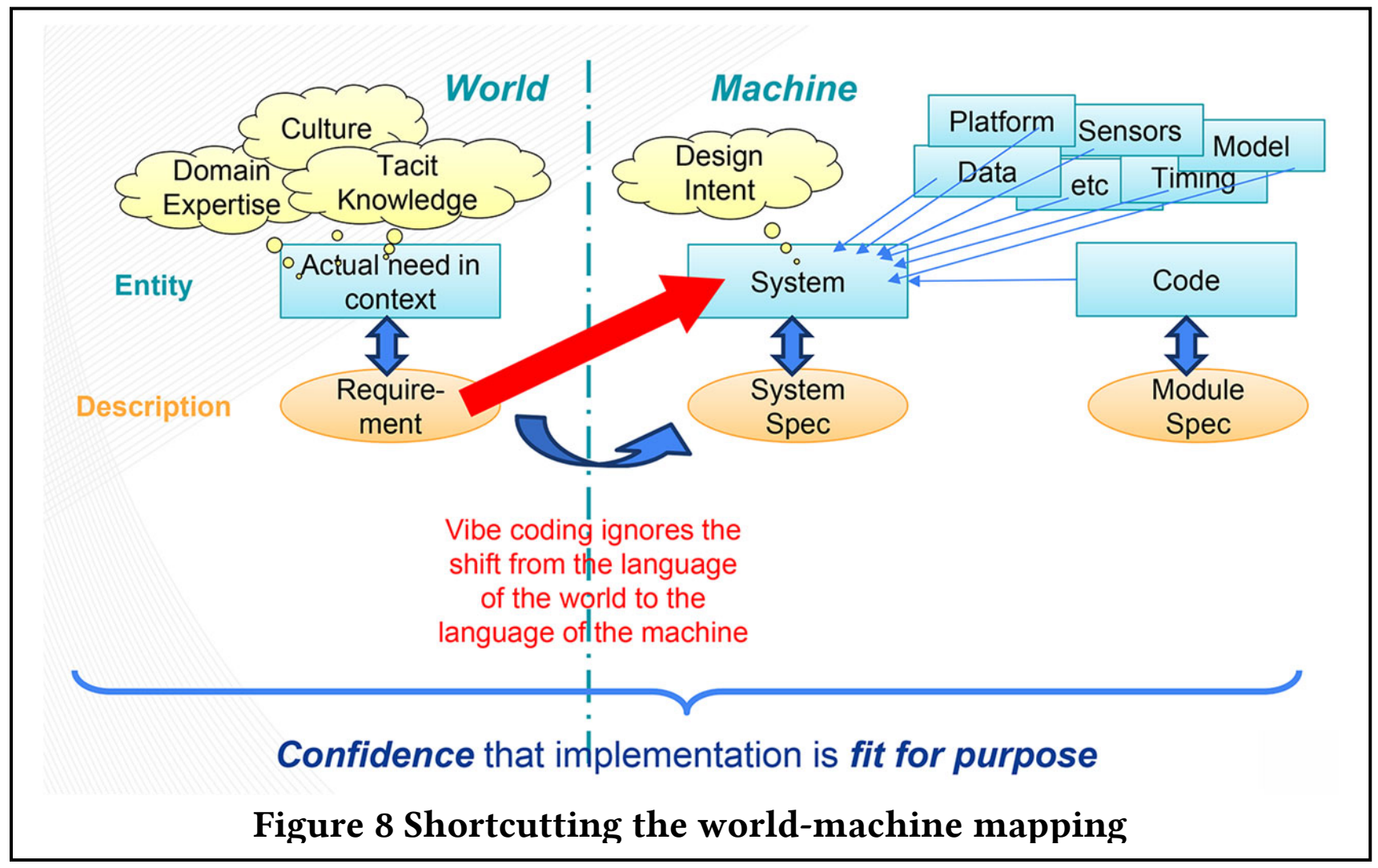

**Figure 8 Shortcutting the world-machine mapping**

Vibe coding seems to be fine for prototypes [33]. However, a prototype is not a product. Prototypes should be (and are in other fields) created to quickly evaluate particular properties and answer particular questions.[7] Too often, software prototypes are treated as early products, even though they lack the infrastructure that makes them reliable, maintainable, and integratable. Further, some properties, such as security, must be engineered into the system from the outset.

**The prototype is not the product**

My concern here, though, is that going directly from a natural language requirement to a software system skips the critical step of showing that the system specification is a faithful rendition of the requirement. Even if the prompt yields a specification, it still depends on an accurate rendition of language of the world into the language of the machine.

The risk of mistranslating from the world language to the machine language is not new. Figure 9 shows a simplified version of an example that Leveson described forty years ago [18]. A computer is used to control a chemical reactor. Sensors report the pressure and temperature in the reactor. The process in the reactor is catalytic: a small amount of a catalyst controls the intensity of the reaction, and controlling the process depends on carefully introducing catalyst in response to the temperature and pressure in the reactor. The requirement for handling anomalies is "If a fault occurs, leave all controlled variables as they were".

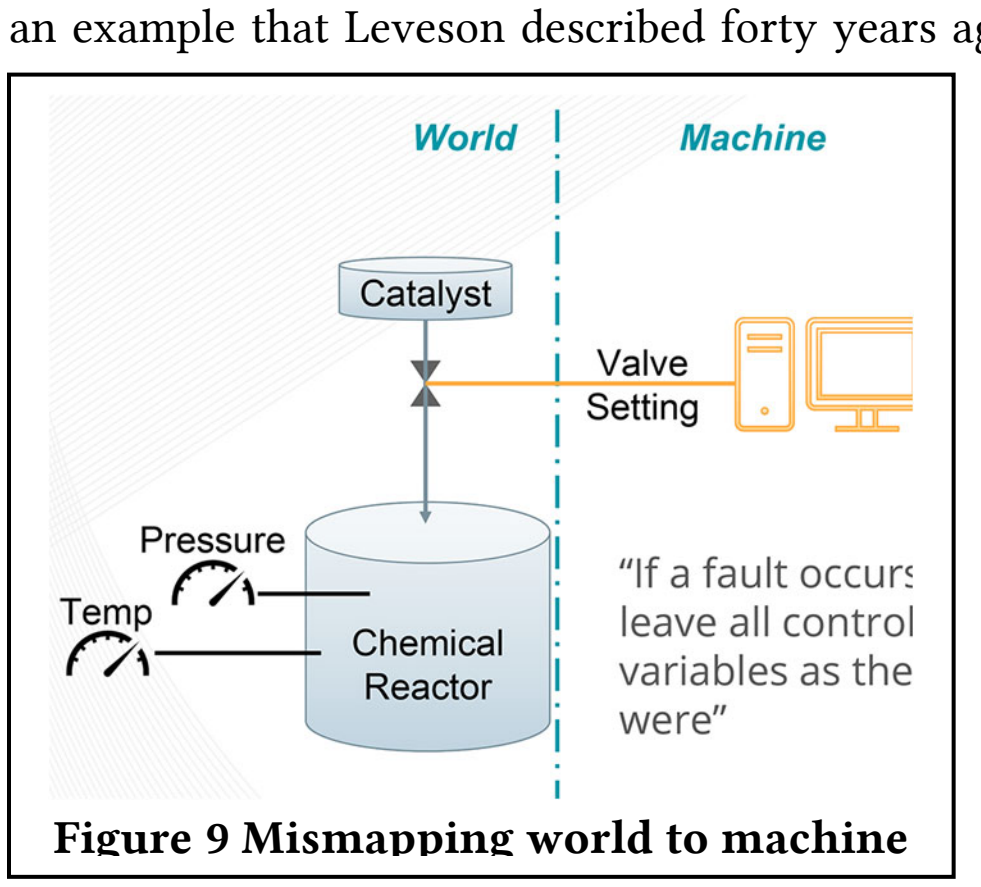

**Figure 9 Mismapping world to machine**

A fault occurred just after the computer controller opened the catalyst valve to add more catalyst. At this point, the difference between the language of the world and the language of the machine becomes critical. The system engineers understood "controlled variables" to be process variables including temperature and pressures. The programmers understood "controlled variables" to mean software values such as the setting of the catalyst valve. The consequence was that the reactor overheated.

The cargo cults described above also shortcut the translation between the desire for cargo and the meaningful actions required to actually bring the flights. The requirement in the world was to summon airplanes with valuable cargo. However their rituals only mimicked the visible activity of the troops—they were informed only by the shadows that the troops' activities had cast on their memories. The rituals were based on that misconception, and they failed.

**Correlation is not causation**

In relying on the digital shadows, the tools may fail to respond to underlying reasons. In modern projects, the oral history may similarly fail to cast meaningful shadows.

## Unsound, statistical basis

The previous two issues involved the language of the world. There is another inherent limitation of LLMs on the machine side, in showing conformance of the system to the system specification. The statistical foundation of LLMS cannot support the hard assurances needed by the safety engineering process.

The V-model visually represents the systems development lifecycle. It depicts the process of going from system requirements to progressively more specific lower-level requirements with rigorous checks at the interfaces, then of progressively checking and integrating the components, with verification and validation at each stage. A supply chain is often involved, with different vendors depending on precise specifications at various stages of the V-model. Koopman [14], in discussing the impact of AI on safety engineering in physical systems, argues that machine learning breaks the V-model because it is probabilistic rather than rigorous.

**You can't get security into a system by sprinkling it like pixie dust on your completed code**

[7] Aircraft designers used to make a balsa model of a fuselage and put it in a wind tunnel to assess its aerodynamic properties. No one ***ever*** stuck a couple of engines on the balsa prototype and called it a product.

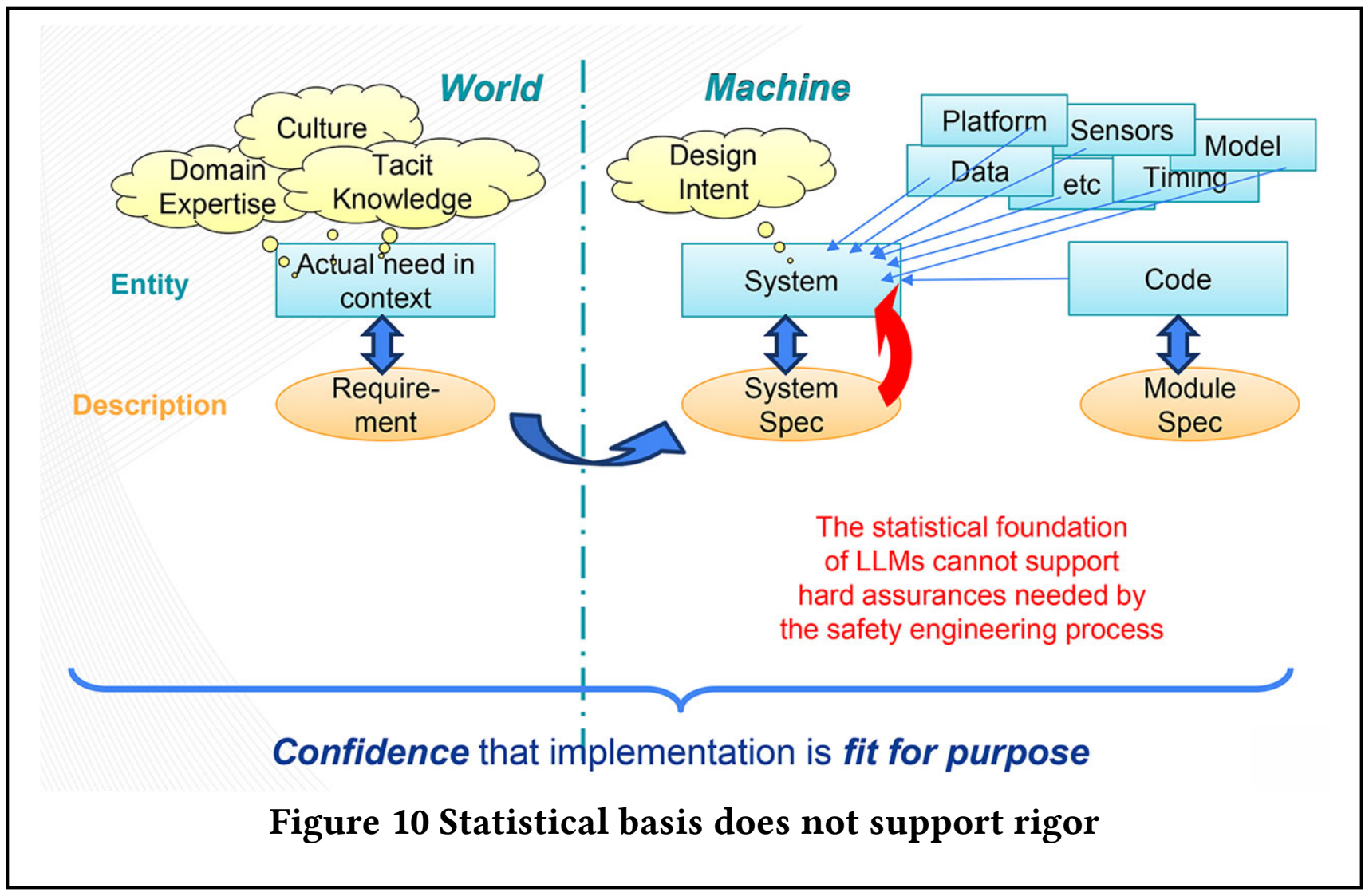

**Figure 10 Statistical basis does not support rigor**

## 5. Specifications and Fitness-for-Purpose

Just as our relation with correctness is complicated, so too is our relation with specifications. Our roots in symbolic, logical, precise reasoning lead us to a view that specifications are formal, static, complete, and rigorous. Just as we recognize that full correctness is often impractical, we must recognize that specifications of this kind are impractical or even impossible. The reasons range from evolution of the problem being solved, to varying levels of rigor in different properties, to incomplete knowledge, to pragmatics about cost-effectiveness.

**Specifications are not formal, static, and complete – they are heterogeneous, evolving, and incomplete**

Specifications and correctness are complementary. Reframing correctness calls for reframing specifications as well. The tension between formality and pragmatics is again strong enough to require a different term. Let's call the expanded framing of specifications "credentials".[8]

Comparing the conventional doctrine on specifications to the practical need shows numerous limitations of specifications:

| **Doctrine on specifications** | **Real-world need: *Credentials*** |
|---|---|
| Sufficient and complete | Incomplete, extensible |
| Static | Incremental, nonmonotonic |
| Correct | Degrees of confidence |
| Functionality (often) | Many properties |
| Homogeneous | Heterogeneous, interdependent |
| Rigorous | Varying degrees of rigor |

Real-world specifications are often incomplete, and they change over time, as do requirements. As new information is acquired, its implications must propagate through the system so that other values that depend on this information can also be updated. Many different properties may affect fitness for purpose, and confidence is not binary but comes on graded scales. Our ability to specify and evaluate different properties has varying degrees of rigor.

Simply tracking properties and their values could be done with association lists, sets of attribute-value pairs. Credentials respond to additional real-world needs by augmenting association lists. The *attribute-value* pairs of a normal association list would suffice to represent the properties of interest and their values, but that is insufficient. We therefore add two additional fields: *credibility*, to record the degree of confidence in the value, and *provenance*, to record the source of the value (e.g., required, empirically measured, proven, received from another component, …) [24][27].

A credential, then, is a set of four-tuples:

***{ < attribute, value, credibility, provenance > }***

The attribute is any of the properties of interest. The value may be of various types, including probability distributions and structured types. The credibility field allows the credential to handle varying degrees of rigor and to rely on techniques from systematic reviews [17] to combine information of varying quality, including qualitative values. The provenance field allows the credential to distinguish required values of attributes from currently-achieved values, and the set representation allows the credential to record (and subsequently combine) different estimates of the values. The provenance field also provides dependency information required to propagate new values as they are established.

For example, a credential might look like

{
<latency, ≤10ms, 90% of the time, requirement>,
<latency, probability distribution, high, model prediction>,
<latency, 8±2 ms, high, benchmark>
<latency, 9±5ms, low, observed>
<latency, risky interactions, medium, human judgment>
}

## 6. Systematic Approach to Assurance

Laying the basis for reframing our approach to correctness introduced a few changes of mindset:

- Reframing "correctness" as establishing our degree of confidence that the software is fit for its purpose
- Recognizing that the criteria for confidence should be determined for each project, setting expected degrees of confidence for each of the important properties
- Accepting the importance of context and that automated tools are working with the digital shadows of reality

[8] This evokes the human act of presenting credentials, but not full CVs, to participate in certain activities. A new ambassador presents formal diplomatic credentials to the head of state in a receiving country. Scuba divers present a scuba certification card in order to get their air tanks filled. A Real ID or equivalent is required to board an airline flight in the US. Your conference badge is a credential. Each of these attests to just a small slice of the owner's life, just certification of the bit that's relevant to the current activity.

- Reframing "specifications" to reflect varied properties and degrees of confidence in each

Let's turn now to how we can be systematic about establishing confidence in fitness for purpose.

### Choosing the properties that matter and the quality required

Fitness for purpose entails confidence in values for select properties. That means seeking sufficient correctness—determining when the system is good enough for its purpose. To do this the designer must.

- Select the important properties for the application, recognizing that specifying and evaluating them has costs
- Determine what degree of confidence is required for each property
- Select the most cost-effective set of development and evaluation techniques

The selected properties usually include quality attributes, not just functionality. They may be properties that can be predicted early in the design or constructed reliably from models as well as properties that can be measured in the running code. Their relative importance may also matter when allocating assurance resources.

**Make deliberate tradeoffs among resources and among quality attributes**

In addition to identifying the properties that matter for the current purpose, the designer needs to determine acceptable values for these properties: speed, latency, probability of failure, accessibility, usability, integrity of system structure, compatibility with other systems, device independence, ... It may also be important to set a level of confidence in the estimation or evaluation of the properties: is full rigor justified, or will empirical data suffice? Is expert judgment a reasonable alternative? Can you just release the software and rely on user feedback?

Predicting or evaluating these properties cost-effectively requires resources for setting up and executing the analyses. Designers must make judgment calls about assurance as well as about implementation.

This calls for an assurance plan as part of the project plan, with resources allocated to assurances that can be analyzed early in design as well as to techniques that analyze the running code.

### Choosing appropriate assurance techniques

SE's complicated relation with correctness includes a lack of consistent, comparable descriptions of assurance techniques. As a result, it's difficult to make principled tradeoffs among assurance techniques that address different properties, at different stages of development, with different degrees of rigor.

Small examples of comparative data have been constructed by practitioners, for practitioners. For example, TMS has a blog with tables that compare assurance techniques [23].

We need a "field guide" to assurance: an open-ended catalog with broad coverage of available techniques and well-structured descriptions that permits comparisons like the ones at [23]. The elements of each description might include, for example,

- Properties analyzed
- Information required for analysis
- Skill required to apply
- Cost
- Scalability
- Coverage
- Confidence in results / degree of rigor
- Stage of development targeted

Some of these lend themselves to quantitative description, others might best be described on ordinal or even nominal scales. The processes for evaluating AI-supported software would fit naturally here, as would risk management techniques appropriate for the statistical basis of AI tools.

If we had such a field guide, the process for assuring that a system was fit for its purpose would be to consult the credential for the properties of interest and their required levels of confidence, then identify the set of assurance techniques that address these properties and can provide the required level of confidence.

**Seek cost-effective solutions to practical problems**

This sets up an optimization problem—more likely a satisficing problem—for choosing the subset of techniques and distribution of resources to answer one of these two questions:

*How much confidence can available resources get me?*
*What resources are needed for my required confidence?*

This would enable us to make deliberate choices about assurance resources and quality attributes, placing AI on the same footing as other tools and techniques.

Any number of techniques might be used for the optimization problem, including applying AI. My purpose here is to frame the question, so I leave this as an open question.

## 7. Reframing Software Assurance for the AI Age

I propose reframing our view of correctness as a unified approach to software assurance that

- judges success as confidence in fitness for purpose
- includes tacit context and the need for human judgment
- recognizes the limitations of relying on digital shadows
- allows for the uncertainties inherent in AI
- integrates evidence with varying degrees of rigor
- supports reasoned allocation of assurance resources

In laying the basis for this reframing I proposed a few changes of mindset.

**Look for dissonance between what everyone says and what everyone does**

- Reframing "correctness" as establishing our degree of confidence that the software is fit for its purpose
- Recognizing that the criteria for confidence should be determined for each project, setting expected degrees of confidence for each of the important properties

- Accepting the importance of context and that automated tools are working with the digital shadows of reality
- Reframing "specifications" to reflect varied properties and degrees of confidence in each

I am indeed suggesting that we back away from the goal of making every program correct. In doing so, I am not suggesting that we abandon our responsibility to correctness. Rather, I am suggesting that we replace our current practice of turning a blind eye to the impracticality of full correctness with recognition of developers' responsibility to determine and achieve an appropriate level of correctness. Further, the traditional position of exclusive attention to correctness fails to take advantage of what we know about risk management and reliability: sometimes it's more cost-effective to detect and fix a problem than it is to prevent it.

**Correctness is a social construct**

**There are two ways to deal with the possibility of Bad Things: prevention and remediation**

My aim is to present an open challenge for the benefit of our community: Let's reframe "correctness" to match real software. The field guide and selection mechanism are open problems, my gift to the SE community. I invite the community to explore this challenge and these ideas.

Perhaps our most important challenge is to learn to reason about the statistical predictions of AI as fluently as we do about symbolic logic. I call on the community to be thoughtfully critical of this new technology and to assimilate it carefully, just as we have assimilated gifts from AI in the past. New technology doesn't necessarily invalidate old theory, though it may require adaptation. Approach AI with this mindset.

**New technology doesn't necessarily invalidate old theory, though it may require adaptation**

In conclusion, we call ourselves engineers. This reframing calls for us to act like engineers, determining how much confidence is needed for fitness of purpose and making tradeoffs to select the suite of assurance techniques that will be most cost-effective.

## ACKNOWLEDGMENTS

I greatly appreciate commentary and advice from Eunsuk Kang, Marian Petre, Margart-Ann Storey, Roy Weil, Bradley Schmerl, Travis Breaux, Christian Kästner, and Wendy Grossman. I thank the FSE chairs for the keynote invitation that instigated working through these ideas. This work was supported by the A.J. Perlis University Professorship in Computer Science at Carnegie Mellon University.

## IMAGE CREDITS

[39] Figures 1, 3, 5, 8 and 10 were drawn by Mary Shaw for this paper.

[40] Figure 2 ecosystem of web development is from [28] (by Mary Shaw) based on https://insights.stackoverflow.com/survey/2019#development-environments-and-tools

[41] Figure 4 Assurance realms is from [28] (by Mary Shaw).

[42] Figure 6 shadow projection of a block is Adobe Stock photo, licensed via Adobe Creative Comons.

[43] Figure 6b and Figure 5 a paper cup and how to fold it are from [4]. Public domain.

[44] Figure 9 is Mary Shaw's simplification of an example in [18].